\documentclass[twocolumn]{aastex631}
\usepackage{silence}
\usepackage{bm}
\usepackage{units}
\usepackage{comment}

\accepted{for publication in The Astrophysical Journal in December 3rd, 2025}

\begin{document}

\title{Electrostatic waves in astrophysical Druyvesteyn plasmas:
       I. Langmuir waves}

\author{Simon Tischmann}
\affiliation{Institut f\"ur Theoretische Physik IV, Ruhr-Universit\"at Bochum, Germany}

\author[0000-0001-5851-7959]{Rudi Gaelzer}
\affiliation{Instituto de Física, Universidade Federal do Rio Grande do Sul, Brazil}

\author[0009-0007-7679-5275]{Dustin Lee Schr\"oder}
\affiliation{Institut f\"ur Theoretische Physik IV, Ruhr-Universit\"at Bochum, Germany}

\author[0000-0002-8508-5466]{Marian Lazar}
\affiliation{Centre for Mathematical Plasma Astrophysics, Department of Mathematics, KU Leuven, Belgium}

\author[0000-0002-9151-5127]{Horst Fichtner}
\affiliation{Institut f\"ur Theoretische Physik IV, Ruhr-Universit\"at Bochum, Germany}

\correspondingauthor{Rudi Gaelzer}
\email{rudi.gaelzer@ufrgs.br}

\begin{abstract}
Plasmas in various astrophysical systems are in non-equilibrium states as evidenced by direct in-situ measurements in the solar wind, solar corona and planetary environments as well as by indirect observations of nonthermal sources of waves and emissions. Specific to observed non-equilibrium plasmas are non-Maxwellian velocity distributions with suprathermal tails, most often described by Kappa (power-law) distributions. In this paper, we introduce an alternative modeling for linear waves in plasmas described by the generalized Druyvesteyn distribution model. This model can reproduce not only high-energy tails, but also low-energy flat-tops of velocity distributions, like those of electrons in interplanetary shocks and the solar transition region. The wave dispersion relation of longitudinal waves is derived in terms of the newly introduced Druyvesteyn dispersion function. The dispersion curves as well as damping rates of high-frequency Langmuir waves are numerically computed for the isotropic case, and their analytical approximations are provided in the limit of weak damping. We thus offer a new tool for modeling longitudinal waves, and in particular Langmuir waves under the specific conditions of Druyvesteyn distributions.
\end{abstract}



\section{Introduction and motivation} 
Many, if not most, astrophysical plasmas are characterized by non-Maxwellian distribution functions, e.g., as measured in situ in the terrestrial magnetosphere \citep{Macek-Wojcik-2023} and the solar wind \citep{Scherer-etal-2022, Abraham-etal-2022}, and are expected for only remotely accessible environments like the solar atmosphere \citep{Dzifcakova-etal-2023}, the interstellar medium \citep{deAvillez-Breitschwerdt-2015}, neutron star magnetospheres \citep{Mousavi-Benacek-2025}, or even quasars \citep{Humphrey-Binette-2014}. The most popular choice for analytically describing non-Maxwellian distributions became, in recent years, the so-called Kappa distribution (for comprehensive overviews see \citet{Livadiotis-2017} and \citet{Lazar-Fichtner-2021}), for which various version are in use, namely standard \citep{Ziebell-Gaelzer-2025}, regularized \citep{Scherer-etal-2017}, generalized \citep{Scherer-etal-2020, Belardinelli-etal-2024, Gaelzer-etal-2024}, and relativistic forms \citep{Han-Thanh-etal-2022}. 

In the context of plasma dispersion theory, there are, however, other non-equilibrium distributions used to describe velocity, momentum, or energy spectra of astrophysical energetic particles. Amongst them we may cite the Weibull distribution \citep[e.g.,][]{Pallocchia-etal-2017}, the Cairns \citep[e.g.,][]{Ayaz-etal-2024} or Vasyliunas-Cairns distribution \citep{Abid-etal-2015}, the (pickup-ion) shell distribution \citep[e.g.,][]{Zank-etal-2001}, the Dory-Guest-Harris distribution \citep[e.g.,][]{Benacek-Karlicky-2019}, or the (generalized) Druyvesteyn distribution \citep{Zaheer-Yoon-2013, Liao-He-2017}. 
The latter is probably the least explored one within an astrophysical context: 
to our knowledge, there are no systematic analyses yet of the properties of observed distributions using the Druyvesteyn model.
This distribution, however, is a potentially useful tool not only to describe suprathermal particle populations but also those being significantly affected by collisions \citep{Liao-He-2017}, and, in particular, the low-energy flat-tops of electron velocity distributions associated with planetary and interplanetary shocks \citep{Feldman-etal-1983a, Feldman-etal-1983b, Zaheer-Yoon-2013, Stasiewicz-2024}.
Such flat-top distributions are also routinely observed in regions where short-range collisions are
not supposed to play an important role, but where the presence of
a strong parallel-aligned component of the electric field induces
the occurrence of intense bidirectional electron beams, such as is
the case with double layers or in the vicinity of the X line in Earth's
magnetotail, where magnetic reconnection takes place \citep{Richard+25/05,Norgren+25/08,Wang+25/09}.

Therefore, it appears worthwhile to systematically study linear plasma waves on the basis of this distribution. 
Here, we begin with electrostatic Langmuir waves, which are of interest in contemporary space physics and astrophysics for, e.g., their excitation and subsequent generation of radio emissions \citep[e.g.,][]{Sauer-etal-2019, Lazar-etal-2022, Lazar-etal-2025}, in the context of turbulence and acceleration or heating of resonant populations \citep{Zaheer-Yoon-2013, Yoon-etal-2024}, and for coherent emission in astrophysical plasmas beyond the heliosphere \citep{Melrose-2017}. 
The case of similarly interesting electrostatic ion-acoustic waves will be considered in a separate analysis.

After the definition of the generalized Druyvesteyn distribution, which we consider to characterize a \textit{Druyvesteyn plasma}, and a 
kinetic derivation
of the corresponding dispersion relation for electrostatic waves (section~2), the resulting dispersion and damping properties of these waves are computed (section~3)\@.
All solutions are obtained for astrophysically relevant parameters both from the analytically derived dispersion relation (formulated in terms of the plasma dispersion function) and by using the state-of-the-art 
\textit{Arbitrary Linear Plasma Solver}
\citep[ALPS, ][]{Verscharen-etal-2018}, which was recently tested and applied in other studies \citep{Schroder-et-al_2025PhPl, Schroder-et-al_2025ApJ}. Finally, all findings are summarized 
(section~4).  

\section{The (Generalized) Druyvesteyn distribution}
\subsection{Definition}
\begin{figure}[ht!]
\includegraphics[width=1\columnwidth]{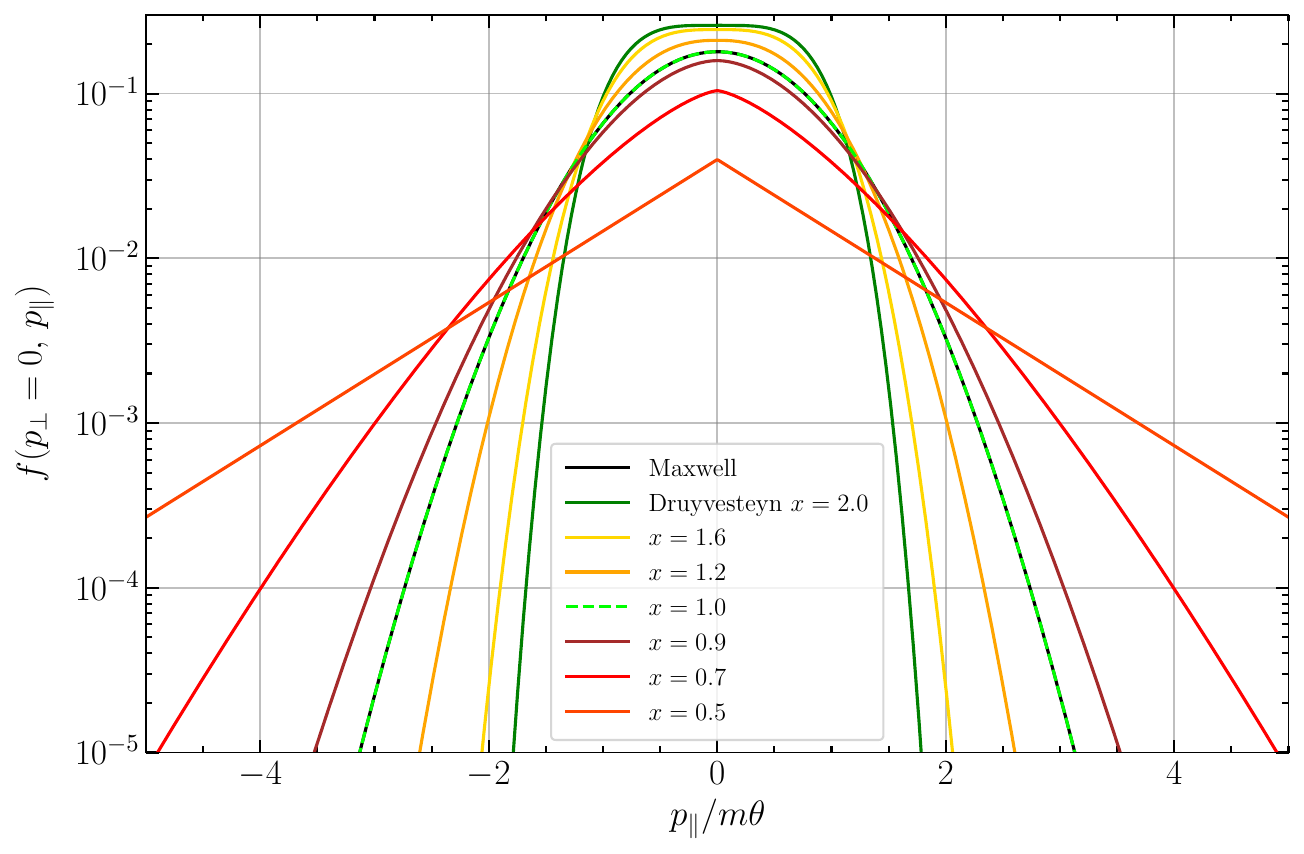}
\caption{%
The generalized Druyvesteyn distribution as defined by Eq.~(\ref{druyvesteyn-dist}), plotted as a function of the normalized particle momentum $p_{\parallel}/(m\theta)$ for $p_{\perp}=0$. The parameter $x$ controls the shape of the distribution, with $x=1$ corresponding to the Maxwellian limit. The plot highlights the gradual deviation from the Maxwellian form with varying $x$.
}
\label{fig:druyvesteyn-dist}
\end{figure}

The original form of the Druyvesteyn distribution was introduced in \citet{Druyvesteyn-1930}, where the influence of energy loss by elastic collisions during electron diffusion was studied. It was generalized in \citet{Behringer-Fantz-1994} to the following form:
\begin{eqnarray}
f_x(E) = A_x \sqrt{E} \exp\left[-\left(\frac{E}{B_x}\right)^x\right],
\label{druyvesteyn-dist}
\end{eqnarray}
with the constants 
\begin{eqnarray}
A_x = \frac{x}{\langle E\rangle^{3/2}_x} \frac{\xi_x^{3/2}}{\Gamma(3/(2x))}
\;\;;\;\;
B_x = \frac{\langle E\rangle_x}{\xi_x}
\end{eqnarray}
and where $\xi_x = \Gamma(5/(2x)) / \Gamma(3/(2x))$ and $x$ will be referred to as the Druyvesteyn parameter. 
It plays an analogous role as the kappa parameter in case of kappa distributions, i.e.\ it parametrizes the family of Druyvesteyn distributions.
The original Druyvesteyn distribution is obtained for $x=2$.
\
The distribution function $f_x(E)$ proposed by \citet{Behringer-Fantz-1994} is valid for any $x > 0$\@.  However, negative values of the parameter $x$ are not physical, since in this case Eq. (\ref{druyvesteyn-dist}) would not be integrable in energy and would not represent a valid probability distribution function.
Figure~\ref{fig:druyvesteyn-dist} illustrates this family of distributions described by Eq. (\ref{druyvesteyn-dist}).

One can clearly observe the formation of a flat-top profile in the low-energy portion of the distribution as $x$ increases $\left(x\gtrsim2\right)$, similar to recently-observed electron distributions in the magnetotail \citep{Wang+25/09}.

\subsection{The Druyvesteyn plasma dispersion function}

Since we are interested in the propagation of longitudinal
waves in an isotropic nonrelativistic plasma, we will first derive
the velocity distribution function (VDF) associated with (\ref{druyvesteyn-dist})\@.
This VDF must be isotropic because it must reduce to the distribution
that describes the thermal equilibrium when $x=1$\@. If $f_{x}\left(E\right)dE$
is the number of particles with energies within the interval $E$
and $E+dE$, the same particles will be contained within a spherical
shell in velocity space with speeds between $v$ and $v+dv$\@.
Hence, given the volume element $4\pi v^{2}dv$, the appropriate velocity
distribution function $f_{x}\left(v\right)$ must be such that
\[
4\pi v^{2}f_{x}\left(v\right)dv=f_{x}\left(E\right)dE.
\]

Given now that $E=\frac{1}{2}mv^{2}$, if $\left\langle E\right\rangle _{x}$
is the mean energy of the particles, then the average value of $v^{2}$
is also given by $\left\langle v^{2}\right\rangle _{x}=2\left\langle E\right\rangle _{x}/m$\@.
Therefore, it follows that $dE=mvdv$ and the appropriate form for
the velocity distribution function is
\[
f_{x}\left(v\right)=\frac{1}{2\pi}\frac{x\xi_{x}^{3/2}}{\Gamma\left(\nicefrac{3}{2x}\right)\left\langle v^{2}\right\rangle _{x}^{3/2}}\exp\left[-\left(\xi_{x}\frac{v^{2}}{\left\langle v^{2}\right\rangle _{x}}\right)^{x}\right].
\]

%
The thermal equilibrium is restored when $x=1$, in which case the
VDF must reduce to the Maxwellian,
\[
f_{1}\left(v\right)=f_{M}\left(v\right)=\frac{e^{-v^{2}/\theta^{2}}}{\pi^{3/2}\theta^{3}},
\]
where $\theta=\sqrt{2T/m}$ is the thermal speed of particles with
mass $m$ and temperature $T$ (in energy units)\@. Therefore, we
must define $\left\langle v^{2}\right\rangle _{x}=\xi_{x}\theta^{2}$,
and we finally obtain the VDF
\begin{equation}
f_{x}\left(v\right)=\frac{xe^{-\left(v^{2}/\theta^{2}\right)^{x}}}{2\pi\Gamma\left(\nicefrac{3}{2x}\right)\theta^{3}}.\label{eq:DRU1:Druyvesteyn_distribution-v}
\end{equation}

The probability distributions (\ref{druyvesteyn-dist}) and (\ref{eq:DRU1:Druyvesteyn_distribution-v}) describe statistical properties of systems with three degrees of freedom.  This is the typical situation for space plasmas and modern spacecraft are equiped to locally measure 3D VDFs.  On the other hand, for some experimental setups, one-dimensional probability distributions of energies or velocities can be relevant.  Such can be the case of the plasma contained in linear confinement machines or for X-ray radiation generated by the incidence of a high-intensity laser on a solid target \citep{Reich+00/05}.

In order to derive the 1D version of the Druyvesteyn velocity distribution function, one has to start not from (\ref{druyvesteyn-dist}) but from the corresponding form of the energy distribution, given by 
\begin{displaymath}
f_{x}^{\mathrm{1D}}\left(E\right) = \frac{x\psi_{x}^{1/2}E^{-1/2}}{\left\langle E\right\rangle _{x}^{1/2}\Gamma\left(\nicefrac{1}{2x}\right)} \exp\left[-\left(\frac{E}{D_{x}}\right)^{x}\right],
\end{displaymath}
where $D_{x} =\psi_{x}^{-1}\left\langle E\right\rangle _{x}$ and $\psi_{x}  = \Gamma\left(\nicefrac{3}{2x}\right)/\Gamma\left(\nicefrac{1}{2x}\right)$\@.  
In the particular case $x = 1$, we recover $f^{\mathrm{1D}}\left(E\right) = E^{-1/2} e^{-E/T}/\sqrt{\pi T}$, which is the energy distribution for a system with one degree of freedom in thermal equilibrium.

Hence, following the same steps we employed starting from (\ref{druyvesteyn-dist}) that ultimately let to the 3D VDF (\ref{eq:DRU1:Druyvesteyn_distribution-v}), we obtain 
\begin{displaymath}
f_{x}^{\mathrm{1D}}\left(v\right)=\frac{xe^{-\left(v^2/\theta^2\right)^x}}
{\Gamma\left(\nicefrac{1}{2x}\right)\theta},
\end{displaymath}
which is the desired form of the Druyvesteyn VDF for a system with one degree of freedom.

Let us now consider an infinite, homogeneous plasma composed by different
particle species/populations, denoted by $a=e,i,\dots$, each described
by the corresponding VDF $f_{a0}\left(\bm{v}\right)$\@.
Let us also consider longitudinal waves with angular frequency $\omega$
propagating parallel to an ambient (homogeneous) magnetic field $\bm{B}_{0}$,
such that the parallel component of the wavevector $\bm{k}$
is $k_{\parallel}=\bm{k}\bm{\cdot}\bm{B}_{0}/B_{0}$\@.
The dispersion relations and the linear damping/growth rates for the
longitudinal normal modes of oscillation are given by the solutions
of the equation \citep{Brambilla98}
\begin{equation}
1+\sum_{a}\chi_{3}^{(a)}=0,\label{eq:Dispersion_equation-gen}
\end{equation}
where
\begin{equation}
\chi_{3}^{(a)}=\frac{\omega_{pa}^{2}}{k_{\parallel}}\int d^{3}v\,\frac{\partial f_{a0}/\partial v_{\parallel}}{\omega-k_{\parallel}v_{\parallel}}
\label{eq:Chi_3-1}
\end{equation}
is the longitudinal component of the susceptibility tensor of the $a$-th plasma species, where now  $\omega_{pa}=\sqrt{4\pi n_{a}q_{a}^{2}/m_{a}}$ is the plasma frequency of the $a$-th species, with $n_{a}$ being its number density, $q_{a}$ the electric charge and $m_{a}$ the mass.

If a given plasma species is Maxwellian, then its partial susceptibility
is given by $\chi_{3,M}^{(a)}=-\omega_{pa}^{2}Z^{\prime}\left(\xi_{a}\right)/k_{\parallel}^{2}\theta_{a}^{2}$,
in terms of the derivative of the well-known Fried \& Conte dispersion
function 
\[
Z\left(\xi\right)=\frac{1}{\sqrt{\pi}}\int_{-\infty}^{\infty}\frac{e^{-t^{2}}dt}{t-\xi}\quad\left(\Im\xi>0\right),
\]
where $\xi=\omega/k_{\parallel}\theta$\@. 

Taking the Maxwellian case as a template, our aim now is to express
the partial susceptibility of a plasma species described by the distribution
function (\ref{eq:DRU1:Druyvesteyn_distribution-v}) in terms of a
new function $Z_{x}\left(\xi\right)$, called \emph{Druyvesteyn plasma
dispersion function}. In order to accomplish this task, we will first
define the nondimensional velocity components $u_{\parallel}=v_{\parallel}/\theta$
and $u_{\perp}=v_{\perp}/\theta$ in cylindrical coordinates and
write in (\ref{eq:Chi_3-1}), 
\[
\int d^{3}v\,\frac{\partial f_{0}/\partial v_{\parallel}}{\omega-k_{\parallel}v_{\parallel}}=-\frac{\theta}{k_{\parallel}}\int d^{3}u\,\frac{\partial f_{0}\left(u_{\perp},u_{\parallel}\right)/\partial u_{\parallel}}{u_{\parallel}-\xi},
\]
where we have taken into account the fact that in a homogeneous plasma
the VDF must by gyrotropic, \emph{i.e.}, $f_{0}\left(\bm{v}\right)=f_{0}\left(v_{\perp},v_{\parallel}\right)$.
Then, after performing an integration by parts in the $u_{\parallel}$
variable, the susceptibility (\ref{eq:Chi_3-1}) can be written as
\begin{equation}
\chi_{3}^{(a)}=-\frac{\omega_{pa}^{2}}{k_{\parallel}^{2}}\theta_{a}\frac{\partial\hphantom{\xi_{a}}}{\partial\xi_{a}}\int d^{3}u\,\frac{f_{0a}\left(u_{\perp},u_{\parallel}\right)}{u_{\parallel}-\xi_{a}}.\label{eq:Chi_3-2}
\end{equation}

Substituting Eq. (\ref{eq:DRU1:Druyvesteyn_distribution-v})
into (\ref{eq:Chi_3-2}) yields
\begin{equation}
I=\int d^{3}u\,\frac{e^{-u^{2x}}}{u_{\parallel}-\xi}=2\pi\int_{-1}^{1}d\mu\int_{0}^{\infty}du\,\frac{u^{2}e^{-u^{2x}}}{u\mu-\xi},\label{eq:Int-I-1}
\end{equation}
where we have introduced the spherical coordinate system $\left(u,\vartheta,\varphi\right)$
and then defined $\mu=\cos\vartheta$. 

The $\mu$-integration in (\ref{eq:Int-I-1}) is
\[
I_{\mu}=\int_{-1}^{1}\frac{d\mu}{u\mu-\xi}=-\frac{2}{u}\tanh^{-1}\left(\frac{u}{\xi}\right).
\]
Hence, we obtain for (\ref{eq:Int-I-1}),
\[
I=-4\pi\int_{0}^{\infty}du\,ue^{-u^{2x}}\tanh^{-1}\left(\frac{u}{\xi}\right),
\]
which will now be integrated by parts, resulting in
\begin{eqnarray*}
I & = & \frac{2\pi}{x}\left[\left.\Gamma\left(x^{-1},u^{2x}\right)\tanh^{-1}\left(\frac{u}{\xi}\right)\right|_{0}^{\infty}\right.\\
 &  & \left.+\xi\int_{0}^{\infty}du\,\frac{\Gamma\left(x^{-1},u^{2x}\right)}{u^{2}-\xi^{2}}\right],
\end{eqnarray*}
where
\[
\Gamma\left(a,z\right)=\int_{z}^{\infty}t^{a-1}e^{-t}dt
\]
is the incomplete gamma function \citep{Paris-Full-NIST10}.

%

It can be verified that, in the last expression for the $I$-integration, the first term inside
the square bracket vanishes. Therefore, the susceptibility component
for an isotropic Druyvesteyn plasma species can be written as 
\begin{equation}
\chi_{3,\mathrm{Dru}}^{(a)}=-\frac{\omega_{pa}^{2}}{k_{\parallel}^{2}\theta_{a}^{2}}Z_{x_{a}}^{\prime}\left(\xi_{a}\right),\label{eq:Chi_3-Druyvesteyn}
\end{equation}
where 
\begin{equation}
Z_{x}\left(\xi\right)=\frac{1}{2\Gamma\left(\nicefrac{3}{2x}\right)}\int_{-\infty}^{\infty}du\,\frac{\Gamma\left(x^{-1},u^{2x}\right)}{u-\xi}\quad\left(\Im\xi>0\right)\label{eq:DRU1:Druyvesteyn_dispersion_function}
\end{equation}
is the Druyvesteyn dispersion function. 
In (\ref{eq:Chi_3-Druyvesteyn}), $Z_{x}^{\prime}(\xi) = dZ_{x}/d\xi$\@.
Since $\Gamma\left(1,z\right)=e^{-z}$,
it can be easily verified that $Z_{1}\left(\xi\right)=Z\left(\xi\right)$,
as expected.

In order to solve the dispersion equation (\ref{eq:Dispersion_equation-gen})
for a plasma containing at least one species described by the Druyvesteyn
distribution, one has to evaluate the susceptibility (\ref{eq:Chi_3-Druyvesteyn}),
which entails the evaluation of the derivative of $Z_{x}\left(\xi\right)$\@.
Derivating (\ref{eq:DRU1:Druyvesteyn_dispersion_function}) \emph{w.r.t.}
$\xi$ and integrating by parts, one obtains
\begin{equation}
Z_{x}^{\prime}\left(\xi\right)=-\frac{x}{\Gamma\left(\nicefrac{3}{2x}\right)}\int_{-\infty}^{\infty}du\,\frac{ue^{-u^{2x}}}{u-\xi}\quad\left(\Im\xi>0\right), \label{eq:DRU1:Druyvesteyn_dispersion_function-derivative}
\end{equation}
which contains as special cases the Maxwellian case ($x=1$, see above) as well as that of the original Druyvesteyn distribution \citep[$x=2$, see][]{Amemiya-2012} and generalizes the results of the latter paper.

Several properties of $Z_{x}^{\prime}\left(\xi\right)$ can be obtained
from (\ref{eq:DRU1:Druyvesteyn_dispersion_function-derivative}),
some of which will be derived in Appendix \ref{Zx-Properties}.

\section{Electrostatic waves in a Druyvesteyn plasma}
\subsection{Numerical solutions of the exact dispersion relation}
In the following, we present solutions of the dispersion relation Eq.(\ref{eq:Dispersion_equation-gen}) for the case of a Druyvesteyn plasma, i.e.\ one where the plasma populations are described by generalized Druyvesteyn distributions Eq.(\ref{druyvesteyn-dist}). This generalizes the results obtained by \citet{Amemiya-2012}, who employed  the original Druyvesteyn distribution (i.e.\ $x=2$) in the context of laboratory plasma applications.

\begin{figure*}[ht!]
\includegraphics[width=1.0\textwidth]{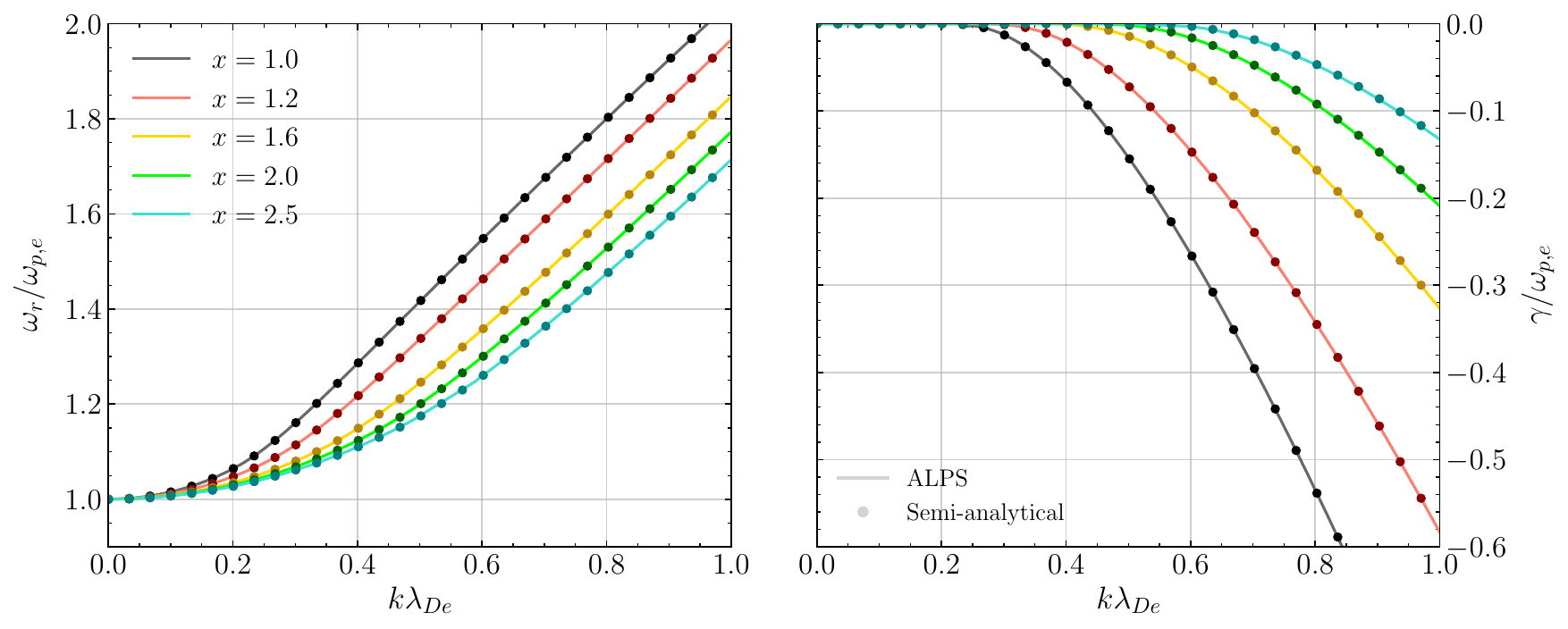}
\caption{Dispersion curves (left) and damping rates (right), both normalized to the electron plasma frequency $\omega_{p,e}$, for a generalized Druyvesteyn distributions with the Druyvesteyn parameter $x > 1$ compared to the Maxwellian ($x=1$). The wave number is normalized using the electron Debye length $\lambda_{De} = \theta d_p/(\sqrt{2} c)$. The solid lines are obtained numerically with the ALPS code directly solving Eq.(\ref{eq:Dispersion_equation-gen}) and the dots are the solutions obtained employing 
the derivative of the Druyvesteyn dispersion function Eq.(\ref{eq:DRU1:Druyvesteyn_dispersion_function-derivative}),
with the properties discussed in Appendix \ref{Zx-Properties}. }
\label{fig:disp_x_gt_1}
\end{figure*}

The dispersion relation Eq.(\ref{eq:Dispersion_equation-gen}) was, on the one hand, solved in its most general form fully numerically with the ALPS code \citep[for a more detailed description of ALPS refer to][]{Schroder-et-al_2025PhPl} by exclusively providing the generalized Druyvesteyn distribution to the code. On the other hand, it was solved after first using the analytical result Eq.(\ref{eq:Chi_3-Druyvesteyn}),
with the expressions for the $Z^\prime_x(\xi)$ function developed in Appendix \ref{Zx-Properties}\@.
Henceforth, we refer to this approach as the semi-analytical solution.
The consistency of the numerical and the semi-analytical solutions corroborates their validity. These solutions were obtained in the wave number interval $k d_p \in [10^{-2},50]$ where $d_p = c/\omega_{p,e}$ denotes the plasma skin depth, i.e.\ the ratio of the speed of light $c$ and the electron plasma frequency $\omega_{p,e}$. 
The choice of this interval makes the results directly comparable to those obtained in \citet{Amemiya-2012} and \citet{Gaelzer-etal-2024}.

The set-up of ALPS requires, first, an input of numerical values of the VDF on a momentum space grid $f(p_{\|}, p_{\perp})$ so that the Druyvesteyn velocity distribution of Eq. (\ref{eq:DRU1:Druyvesteyn_distribution-v}) is implemented in the form
%
\begin{eqnarray*}
f_x(p) = f_x(p_\|,p_\perp) = \frac{x/(2\pi)}{ \Gamma(\nicefrac{3}{2x}) m^3 \theta^3}
         \exp\left[-\left(\frac{p^2_{\|}+p^2_\perp}{m^2\theta^2}\right)^{\!\!x}\right]
\end{eqnarray*}
to generate an ASCII table containing the values of $f(p_{\|}, p_{\perp})$\@. 
Second, an Alfv\'en speed has to be defined. It is taken to be $v_A = 10^{-2} c.$ in the present case, namely $\theta = 3 v_A$. The wave number is normalized using the electron Debye length $\lambda_{De} = \theta d_p/(\sqrt{2} c)$. Starting the iteration in ALPS with $\omega/\omega_{p,e} = 1$ and $\gamma/\omega_{p,e} = -6\cdot 10^{-4}$ and using a resolution of 128 $k$-intervals, results in the dispersion curves and damping rates shown in Fig.~\ref{fig:disp_x_gt_1}. 

The figure reveals that  (i) compared to a Maxwellian plasma ($x=1$), the dispersion curves in a Druyvesteyn plasma with $x>1$ stay longer at frequencies closer to the plasma frequency so that the turn-over into the 'thermal' branch occurs at higher wave number, and (ii) that the damping decreases with increasing $x$. The results for the original Druyvesteyn distribution ($x=2$ are in agreement with those found by \citet{Amemiya-2012}. 

\begin{figure*}[ht!]
\includegraphics[width=1.0\textwidth]{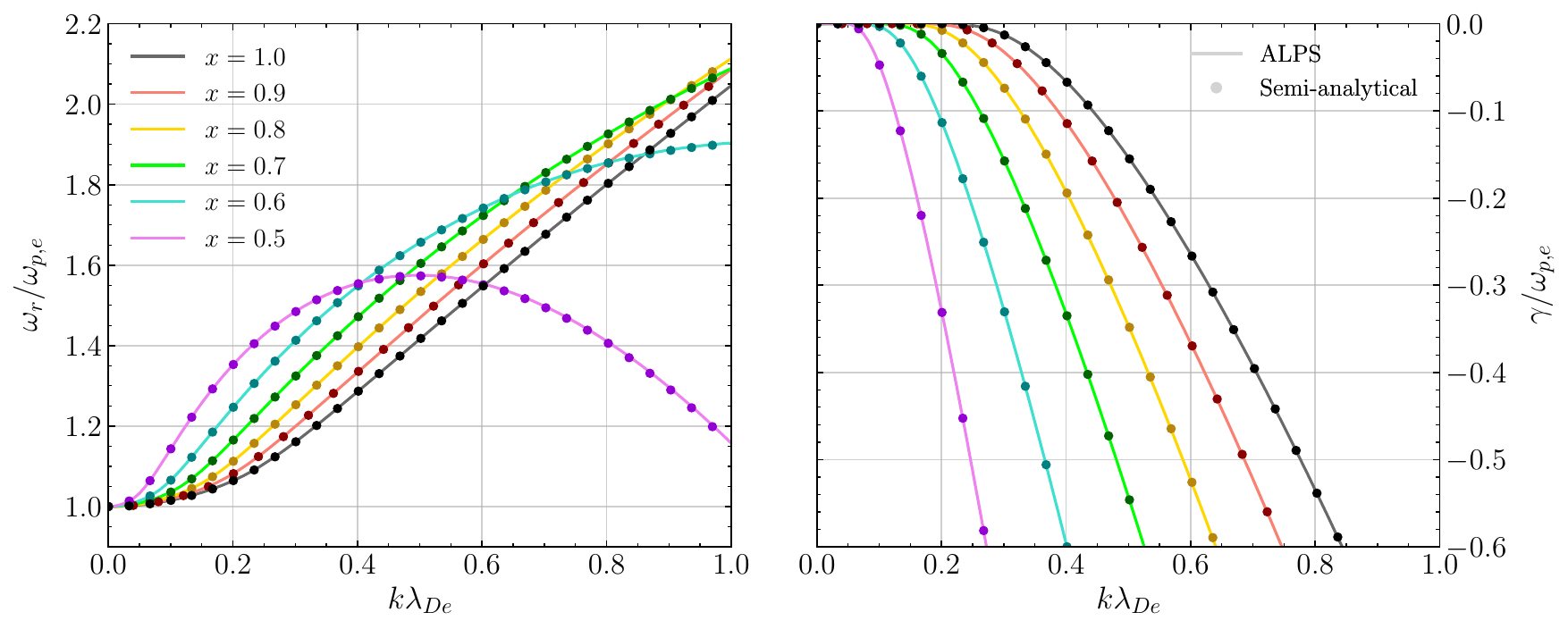}
\caption{
        Dispersion curves (left) and damping rates (right), both normalized to the electron plasma frequency $\omega_{p,e}$, for a generalized Druyvesteyn distributions with the Druyvesteyn parameter $x < 1$ compared to the Maxwellian ($x=1$). The wave number is normalized using the electron Debye length $\lambda_{De} = \theta d_p/(\sqrt{2} c)$. The solid lines are obtained numerically with the ALPS code directly solving Eq.(\ref{eq:Dispersion_equation-gen}) and the dots are the solutions obtained employing 
the derivative of the Druyvesteyn dispersion function Eq.(\ref{eq:DRU1:Druyvesteyn_dispersion_function-derivative}),
with the properties discussed in Appendix \ref{Zx-Properties}. }
\label{fig:disp_x_lt_1}
\end{figure*}

As can be expected, for a Druyvesteyn plasma with $x < 1$ this behaviour is reversed, see Fig.~\ref{fig:disp_x_lt_1}. With decreasing Druyvesteyn parameter $x$, the frequency is increasing for given wave number if the latter is sufficiently low and the damping is increasing. Interestingly, however, the dispersion curves do not turn asymptotically in the thermal branch with constant slope, but turn back to lower frequencies. This effect is more pronounced with decreasing $x$, i.e.\ with an increasing amount of suprathermal particles (see Fig.~\ref{druyvesteyn-dist}).  

The occurrence of such anomalous dispersion, i.e.\ dispersion curves with negative slope ($d\omega_r/dk < 0$) in the presence of a significant amount of suprathermal particles, here electrons, has been observed before in studies of Langmuir waves. For example, \citet{Timofeev-2013} has investigated the dispersion and Landau damping of Langmuir waves in a non-Maxwellian plasma characterized by a truncated power-law momentum distribution function. He found that oscillations with shorter wavelength are subject to strong Landau damping and exhibit anomalous dispersion (see Fig.~3 in that paper). Instead of a hard truncation of a power-law, a physically better motivated approach assumes an exponential cut-off of a power law, which led in the case of the frequently employed kappa distributions to the definition of \textit{regularized} kappa distributions \citep{Scherer-etal-2017, Lazar-Fichtner-2021}. Using the latter, \citet{Gaelzer-etal-2024} have studied Langmuir waves and also found anomalous dispersion accompanied by strong damping in the presence of strongly suprathermal electrons (see their Fig.~4). Of course, anomalous dispersion and the associated phenomenon of a negative group velocity are not unique to Langmuir waves but are studied in broader contexts like electron-beam plasmas \citep{Sukhomlinov-etal-2021}, dusty plasmas \citep{Motcheyo-etal-2018, deToni+22/11}, quantum plasmas \citep{Haas-etal-2012}, or laboratory plasmas \citep{Coppi-etal-1990}.

\subsection{Analytical approximation in the limit of weak damping}



The behavior of the dispersion relation of Langmuir waves and the associated damping rate observed in Figs. \ref{fig:disp_x_gt_1} and \ref{fig:disp_x_lt_1} as the
parameter $x$ varies can be better understood when one derives approximate analytical expressions for these physical properties.

Returning to the dispersion equation (\ref{eq:Dispersion_equation-gen})
for longitudinal waves, we will consider the propagation of Langmuir
waves in an electron-ion plasma. Since Langmuir waves have high frequency,
usually only the contribution of electrons is included. If the electrons
are described by the Druyvesteyn distribution (\ref{eq:DRU1:Druyvesteyn_distribution-v}),
the partial susceptibility in this case is given by (\ref{eq:Chi_3-Druyvesteyn})
and the dispersion equation for Langmuir waves becomes
\begin{equation}
\Lambda\left(k_{\parallel},\omega\right)=1-\frac{\omega_{pe}^{2}}{k_{\parallel}^{2}\theta_{e}^{2}}Z_{x_{e}}^{\prime}\left(\xi_{e}\right)=0,\label{eq:DRU1:DELD-Langmuir}
\end{equation}
where $\xi_{e}=\omega/k_{\parallel}\theta_{e}$\@.

If we write the solutions of (\ref{eq:DRU1:DELD-Langmuir}) as $\omega\left(k_{\parallel}\right)=\omega_{r}\left(k_{\parallel}\right)+i\gamma\left(k_{\parallel}\right)$,
approximate analytical expressions for $\omega_{r}\left(k_{\parallel}\right)$
and $\gamma\left(k_{\parallel}\right)$ can be derived from the weak
absorption assumption,
which is valid when $\left|\gamma\right|\ll\omega_{r}$
and $\left|\gamma\partial\Lambda_{i}/\partial\omega_{r}\right|\ll\left|\Lambda_{r}\left(\omega_{r}\right)\right|$,
where we have also written $\Lambda\left(\omega_{r}\right)=\Lambda_{r}\left(\omega_{r}\right)+i\Lambda_{i}\left(\omega_{r}\right)$.

According to equation (\ref{eq:DRU1:DDFP-Analytic_continuation}), we can split
(\ref{eq:DRU1:DELD-Langmuir}) as 
\begin{eqnarray*}
\Lambda_{r}\left(k_{\parallel},\omega_{r}\right) & = & 1-\frac{\omega_{pe}^{2}}{k_{\parallel}^{2}\theta_{e}^{2}}Z_{x,NC}^{\prime}\left(\xi_r\right)\\
\Lambda_{i}\left(k_{\parallel},\omega_{r}\right) & = &
\frac{\pi x}{\Gamma\left(\nicefrac{3}{2x}\right)}
\frac{\omega_{pe}^{2}}{k_{\parallel}^{2}\theta_{e}^{2}}
\xi_r e^{-\xi_r^{2x}},
\end{eqnarray*}
where $\xi_r = \omega_r/k_\parallel \theta_e$\@.
Hence, according to the weak absorption approximation \citep{KrallTrivelpiece86}, the dispersion
relation $\omega_{r}\left(k_{\parallel}\right)$ will be determined
by the solution of $\Lambda_{r}\left(k_{\parallel},\omega_{r}\right)=0$,
whereas the absorption rate is given by 
\begin{equation}
\gamma(k_\parallel) = -\frac{\Lambda_{i}\left(k_{\parallel},\omega_{r}\right)}{\partial\Lambda_{r}/\partial\omega_{r}}.\label{eq:DRU1:Weak_resonance-approximation}
\end{equation}

Since Langmuir waves are assumed to be fast waves, \emph{i.e.}, $\omega_{r}/k_{\parallel}\gg\theta_{e}$,
we can evaluate $\Lambda_{r}\left(k_{\parallel},\omega_{r}\right)$
using the asymptotic expansion (\ref{eq:DRU1:DDF-asymptotic}) for
$Z_{x,NC}^{\prime}$\@. Hence,
\[
Z_{x,NC}^{\prime}\left(\frac{\omega_{r}}{k_{\parallel}\theta_{e}}\right)\approx\left(\frac{k_{\parallel}\theta_{e}}{\omega_{r}}\right)^{2}\left[1+\frac{\Gamma\left(\nicefrac{5}{2x}\right)}{\Gamma\left(\nicefrac{3}{2x}\right)}\left(\frac{k_{\parallel}\theta_{e}}{\omega_{r}}\right)^{2}\right],
\]
and we have 
\[
\Lambda_{r}\left(k_{\parallel},\omega_{r}\right)\approx1-\frac{\omega_{pe}^{2}}{\omega_{r}^{2}}\left[1+\frac{\Gamma\left(\nicefrac{5}{2x}\right)}{\Gamma\left(\nicefrac{3}{2x}\right)}\frac{k_{\parallel}^{2}\theta_{e}^{2}}{\omega_{r}^{2}}\right]=0,
\]
from which one easily obtains
\begin{equation}
\omega_{x,DBG}\left(k_{\parallel}\right)=\omega_{pe}\sqrt{1+\frac{\Gamma\left(\nicefrac{5}{2x}\right)}{\Gamma\left(\nicefrac{3}{2x}\right)}\frac{k_{\parallel}^{2}\theta_{e}^{2}}{\omega_{pe}^{2}}},\label{eq:DRU1:Druvesteyn-Bohm-Gross-dispersion_relation}
\end{equation}
where $\omega_{x,DBG}\left(k_{\parallel}\right)$ is henceforth called
the \emph{Druyvesteyn-Bohm-Gross dispersion relation}.

For the Druyvesteyn parameter $x=1$, i.e.\ the Maxwellian case, we obtain 
\[
\omega_{1,DBG}\left(k_{\parallel}\right)\approx\omega_{pe}\sqrt{1+\frac{3}{2}\frac{k_{\parallel}^{2}\theta_{e}^{2}}{\omega_{pe}^{2}}}=\omega_{BG}\left(k_{\parallel}\right),
\]
where $\omega_{BG}\left(k_{\parallel}\right)$ is the well-known Bohm-Gross
dispersion relation.

Now, for the evaluation of the absorption rate, we have, to lowest
order, 
\[
\frac{\partial\Lambda_{r}}{\partial\omega_{r}}\approx2\frac{\omega_{pe}^{2}}{\omega_{r}^{3}}\approx\frac{2}{\omega_{pe}},
\]
and from (\ref{eq:DRU1:Weak_resonance-approximation}) one obtains
\begin{equation}
\frac{\gamma_{x}\left(k_{\parallel}\right)}{\omega_{x,DBG}}=-\frac{\pi x}{2\Gamma\left(\nicefrac{3}{2x}\right)}\left(\frac{\omega_{pe}}{k_{\parallel}\theta_{e}}\right)^{3}\exp\left[-\left(\frac{\omega_{x,DBG}^{2}}{k_{\parallel}^{2}\theta_{e}^{2}}\right)^{x}\right],\label{eq:Druyvesteyn-weak_absorption}
\end{equation}
which is the weak absorption rate of Langmuir waves in an electron-Druyvesteyn
plasma.

Once again, 
\[
\frac{\gamma_{1}}{\omega_{BG}}=-\sqrt{\pi}\left(\frac{\omega_{pe}}{k_{\parallel}\theta_{e}}\right)^{3}e^{-\omega_{BG}^{2}\left(k_{\parallel}\right)/k_{\parallel}^{2}\theta_{e}^{2}},
\]
which is the standard expression for the (weak) absorption rate in
a Maxwellian plasma.

\begin{figure*}
\begin{minipage}[t]{0.49\textwidth}%
\includegraphics[width=1\linewidth]{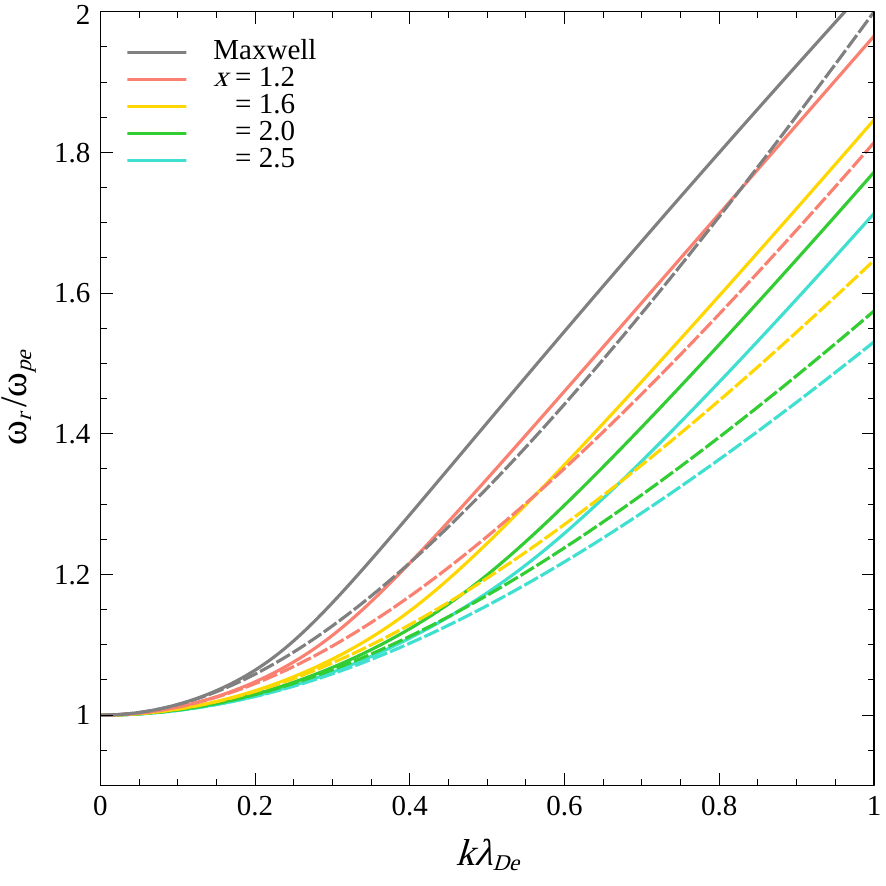}%
\end{minipage}\hfill{}%
\begin{minipage}[t]{0.49\textwidth}%
\includegraphics[width=1\linewidth]{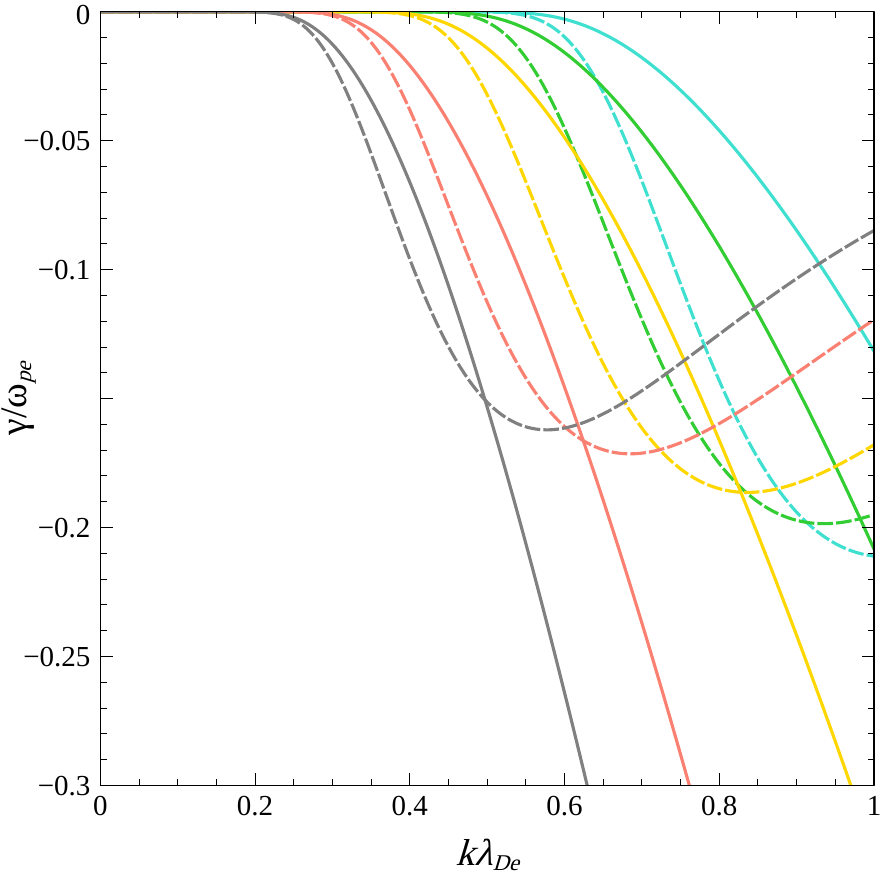}%
\end{minipage}

\begin{minipage}[t]{0.49\textwidth}%
\includegraphics[width=1\linewidth]{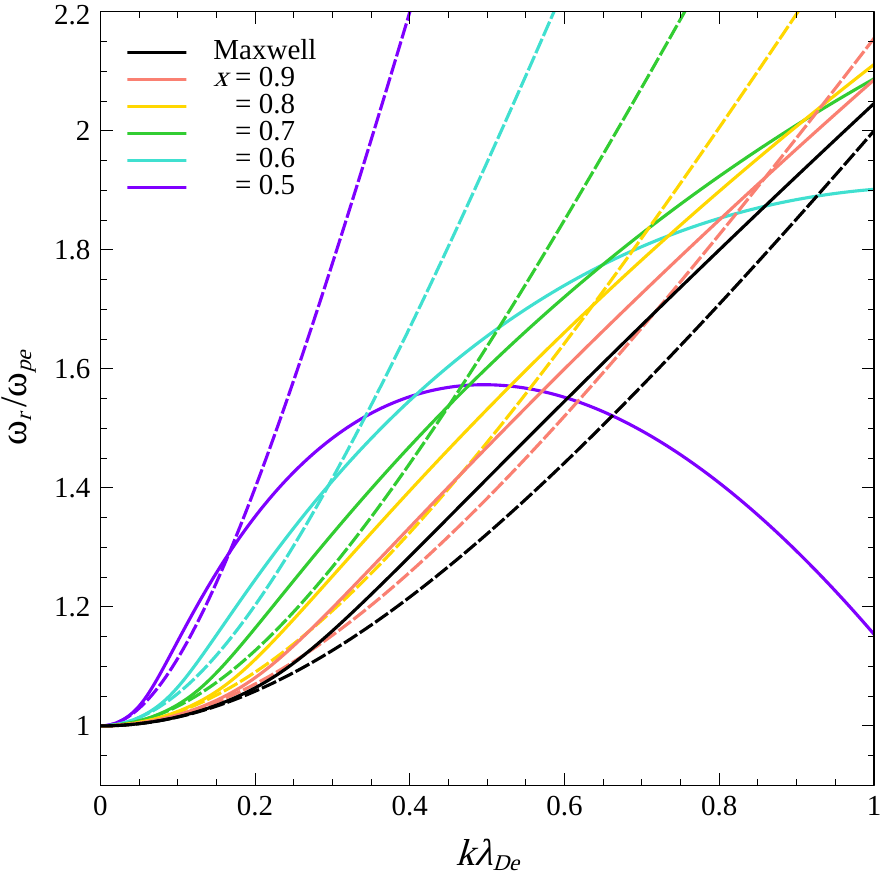}%
\end{minipage}\hfill{}%
\begin{minipage}[t]{0.49\textwidth}%
\includegraphics[width=1\linewidth]{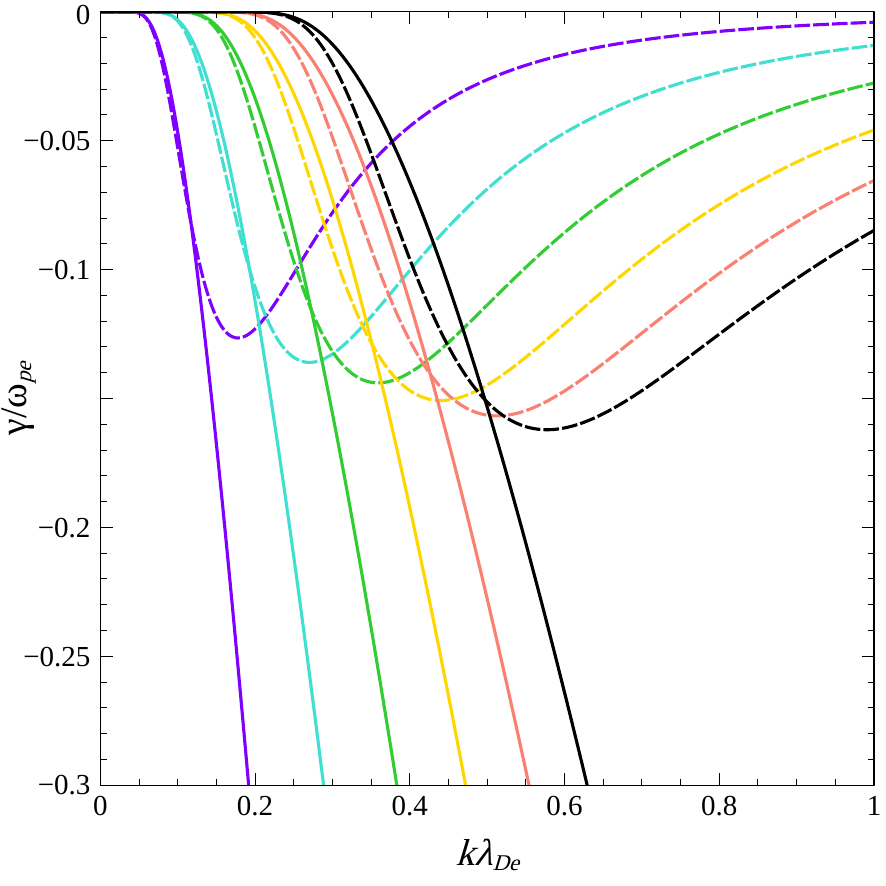}%
\end{minipage}

\caption{Comparison of the approximate expressions (Eqs. \ref{eq:DRU1:Druvesteyn-Bohm-Gross-dispersion_relation}
and \ref{eq:Druyvesteyn-weak_absorption}) (dashed lines) with the
full numerical solution of Eq. (\ref{eq:DRU1:DELD-Langmuir}) (solid
lines)\@. The plots show dispersion relation curves (left column)
and damping rates (right column) for a Druyvesteyn plasma with $x>1$
(upper row) and $x<1$ (lower row)\@.}\label{fig:DRU1:Tischmann-2024-1}
\end{figure*}

Figure \ref{fig:DRU1:Tischmann-2024-1} shows plots of the full numerical
solutions of the dispersion equation (\ref{eq:DRU1:DELD-Langmuir})
in continuous lines, which were obtained with the expressions for
$Z_{x}^{\prime}\left(\xi\right)$ derived in Appendix \ref{Zx-Properties}\@.
Also shown are plots of the approximate dispersion relation (\ref{eq:DRU1:Druvesteyn-Bohm-Gross-dispersion_relation})
and of the absorption rate (\ref{eq:Druyvesteyn-weak_absorption}),
in dashed lines. The top plots are for $x>1$ whereas the bottom plots
display cases with $x<1$\@. In both cases, plots for a Maxwellian
electron plasma $\left(x=1\right)$ are included for reference.
The continuous lines reproduce the plots displayed in Figs. \ref{fig:disp_x_gt_1} and \ref{fig:disp_x_lt_1}.

As it happens already with a Maxwellian plasma, the approximate expressions
have limited validity.
The dispersion relations are accurate only for $k_{\parallel}\lambda_{De}\lesssim 0.4$
for the cases with $x > 1$, and for even smaller spectral ranges when $x < 1$\@.
For low wavenumbers, the approximate expressions predict lower frequencies
than the full numerical solutions, but both remain relatively close until they eventually cross.  For a Maxwellian plasma, the crossing occurs at $k_\parallel \lambda_{De} \approx 1.2$ (not shown), but the crossing point moves to higher wavenumbers as the parameter $x$ grows.
After the crossing point, the approximate solution diverges from the full numerical solution.
%

The behaviour is the opposite when $x < 1$\@.  In this case, the crossing point moves towards zero, as can be seen in the bottom panels of Fig. \ref{fig:DRU1:Tischmann-2024-1}\@.  In particular, when $x = 0.5$, the crossing occurs at $k_\parallel \lambda_{De} \approx 0.17$\@.

Since the approximate solutions are derived with the assumption that
Langmuir waves are fast $(\xi_r \gg 1)$, the analytical dispersion relations are
accurate when the wave resonates with electrons at the very tail of
the distribution. Consequently, the approximation is better when the
distribution is super-Maxwellian ($x>1$, see Fig. \ref{fig:druyvesteyn-dist})\@.
Conversely, in the sub-Maxwellian case ($x<1$, bottom panels of Fig.
\ref{fig:DRU1:Tischmann-2024-1}), the analytical dispersion relations
become less accurate sooner as $x$ decreases, because the distribution
becomes flatter and electrons from the core start to resonate with the waves.
Indeed, it is evident that after the
analytical and numerical curves for $\omega_{r}\left(k_{\parallel}\right)$
cross, they diverge at a rate faster than linear. In fact, when $x<1$,
the full dispersion relations display an anomalous behavior, as is
evident in the case $x = 0.5$, where $\omega_{r}$ starts
to decrease after $k_{\parallel}\lambda_{De}\simeq0.4$\@. 
This anomalous behavior was already commented upon in the discussion regarding Figs. \ref{fig:disp_x_gt_1} and \ref{fig:disp_x_lt_1}.

The divergence between the analytical approximation and the full numerical solution of the dispersion relation after the crossing point happens even with a Maxwellian plasma and is due to the fact that for high wavenumbers the wave is strongly damped and the initial assumption of weak resonance $(|\gamma| \ll \omega_r)$ is no longer valid.
However, within their validity range, the analytical expressions provide a simple physical interpretation for the salient aspects of the dispersion relations and damping rates.
Regarding the dispersion relations, the results show that for super-Maxwellian plasmas $(x > 1)$ the frequency of the Langmuir waves at a given wavenumber reduces as $x$ increases, whereas for sub-Maxwellian plasmas the frequency increases as $x$ decreases.

\begin{figure}[ht!]
\includegraphics[width=1\columnwidth]{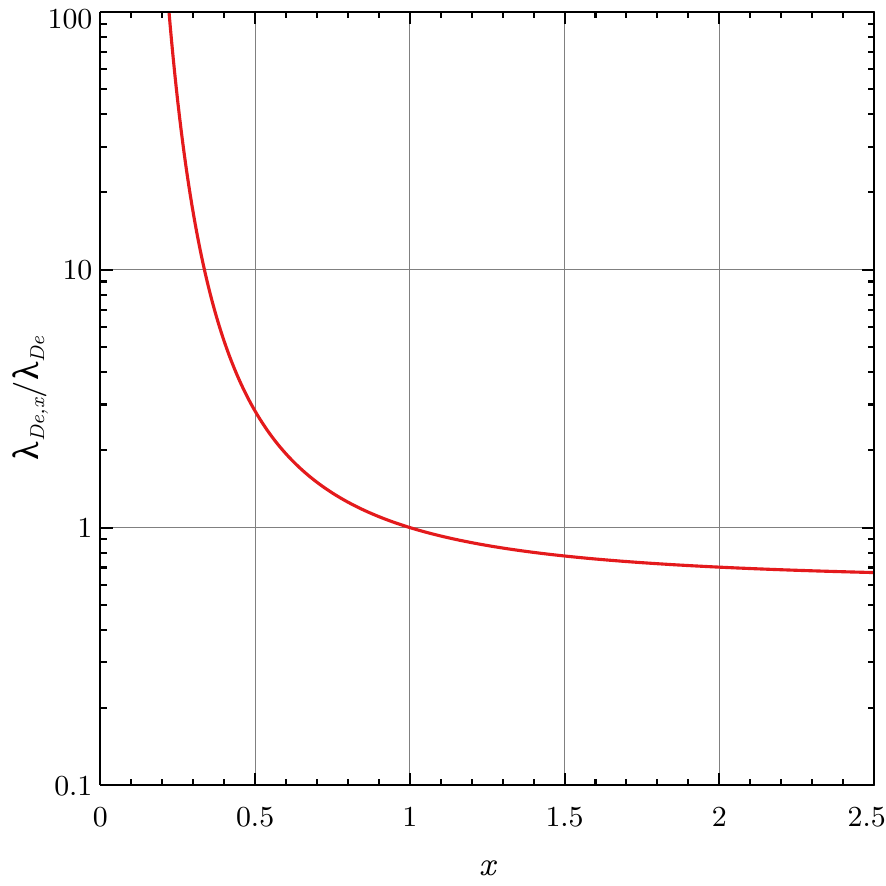}
\caption{
The effective Debye length as a function of $x$.
}
\label{fig:DRU1:Druyvesteyn_Debye}
\end{figure}

Returning to the Druyvesteyn-Bohm-Gross dispersion relation (\ref{eq:DRU1:Druvesteyn-Bohm-Gross-dispersion_relation}), we can define an effective electron Debye length $\lambda_{De,x}$ for a Druyvesteyn plasma as
\begin{equation}
\lambda_{De,x} = g_x \lambda_{De}, \text{ where }
g_x = \sqrt{\frac{2}{3}\frac{\Gamma\left(\nicefrac{5}{2x}\right)}{\Gamma\left(\nicefrac{3}{2x}\right)}}
\lambda_{De}, \label{eq:DRU1:Effective-Debye_length}
\end{equation}
and where $\lambda_{De} = \theta_e/\sqrt{2}\omega_{pe}$ is the Debye length for a Max\-wel\-li\-an plasma.
In this way, (\ref{eq:DRU1:Druvesteyn-Bohm-Gross-dispersion_relation}) can be written as
$\omega_{x}\left(k_{\parallel}\right)=\omega_{pe}\sqrt{1+3\lambda_{De,x}^{2}k_{\parallel}^{2}}$, formally identical to the Bohm-Gross dispersion relation $\omega_{BG}(k_\parallel)$\@.
Figure \ref{fig:DRU1:Druyvesteyn_Debye} shows the plot of $\lambda_{De,x}$ as a function of $x$\@. One observes that $\lambda_{De,x} \lessgtr \lambda_{De}$  for $x \gtrless 1$.

As a consequence of the variation of the effective Debye length with $x$, a Langmuir wave propagating in a super-Maxwellian plasma will have a given frequency $\omega_r > \omega_{pe}$ at a higher wavenumber in comparison with the same wave propagating in a thermal plasma, with the result that the whole dispersion curve displays a smaller variation with $k_\parallel$ and becomes flatter as $x$ increases.  This behavior remains even beyond the crossing point.
Conversely, in a sub-Maxwellian plasma the wave frequency changes faster with $k_\parallel$ and the dispersion curve results steeper than in a thermal plasma.  In this case, however, the anomalous behavior beyond the crossing point results in diminishing frequencies when the wave-particle resonance is strong.

This behavior will have an important effect on the radiation flux in a Druyvesteyn plasma. Since the energy flux convectively transported by oscillations within a plasma is assumed, on the first order, to occur at the group velocity, the small resonance approximation predicts that $v_{g,x} = (1 - \omega_{pe}^{2}/\omega_{x}^{2}) v_{\phi,x}$,
where $v_{\phi,x} = \omega_x/k_\parallel$ is the phase velocity. Comparing with the group velocity in a thermal plasma $(v_g)$, we observe that
\begin{displaymath}
\frac{v_{g,x}}{v_{g}} \approx \left(\frac{\lambda_{De,x}}{\lambda_{De}}\right)^{2},
\end{displaymath}
within the validity of the approximation.  Hence, the energy transported by Langmuir waves will propagate slower in a super-Maxwellian plasma and faster in a sub-Maxwellian plasma.
In the anomalous dispersion region, the behavior of the group velocity radically changes and it can even change sign.  However, in this case the wave is strongly absorbed and can only propragate in short distances anyway.

Let us now focus on the behavior of absorption rates, displayed by the RHS panels of Fig. \ref{fig:DRU1:Tischmann-2024-1}\@.  We once again point out that the analytical expression (\ref{eq:Druyvesteyn-weak_absorption}) is only valid in a finite spectral range with low wavenumbers.
Notice that even when $x=1$, the analytical expression
for $\gamma_{x}\left(k_{\parallel}\right)$ predicts that the absorption
rates are finite throughout the spectral range, whereas, in reality,
$\left|\gamma_{x}\left(k_{\parallel}\right)\right|$ apparently always
increases with $k_{\parallel}$, becoming eventually comparable to $\omega_r$.

As is well known, within the weak resonance assumption, the damping rate of longitudinal waves propagating in a plasma with an arbitrary number of species or populations is given by
\begin{displaymath}
\gamma = \pi\left(\frac{\partial\Lambda_{r}}{\partial\omega_{r}}\right)^{-1}\sum_{a}\frac{\omega_{pa}^{2}}{k_{\parallel}^{2}}\left.\frac{dF_{a0}}{dv_{\parallel}}\right|_{v_{\parallel}-\omega_{r}/k_{\parallel}},
\end{displaymath}
where $F_{a0}\left(v_{\parallel}\right)=2\pi\int_{0}^{\infty}dv_{\perp}\,v_{\perp}f_{a0}\left(v_{\parallel},v_{\perp}\right)$ is the integrated (in the perpendicular velocity direction) VDF for species $a$\@.  That is, the damping (or growth) rate is proportional to the derivatives of the VDFs at the resonant velocity.

In a Druyvesteyn plasma, the corresponding expression for the integrated VDF, $F_x = F_x(v_\parallel)$, can be obtained from (\ref{eq:DRU1:Druyvesteyn_distribution-v}), but this is not necessary.  For a semiquantitative discussion, it suffices to observe that  $\gamma \propto \Lambda_i(k_\parallel,\omega_r) \propto Z^\prime_{x,C}(\xi_r)$, according to (\ref{eq:DRU1:Weak_resonance-approximation}) and (\ref{eq:DRU1:DDFP-Analytic_continuation})\@.
Hence, we can understand the behavior of the damping rate by looking at the continued part of the derivative of the dispersion function at the resonant velocity, which must be such that $\xi_r \gg 1$.

\begin{figure}[ht!]
\includegraphics[width=1\columnwidth]{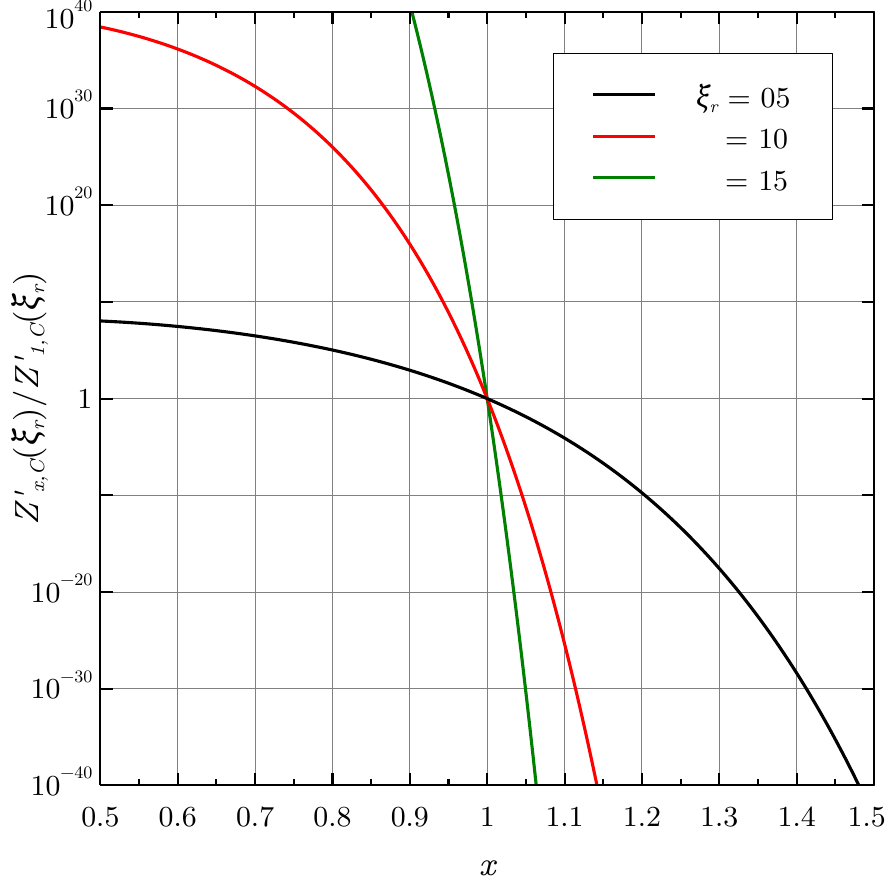}
\caption{
The dependence of $Z^\prime_{x,C}(\xi_r)$ as a function of $x$.
}
\label{fig:DRU1:Druyvesteyn_Zp_ratios}
\end{figure}

Figure \ref{fig:DRU1:Druyvesteyn_Zp_ratios} shows how the function $Z^\prime_{x,C}(\xi_r)$  behaves as a function of the parameter $x$, as compared with the Maxwellian case $Z^\prime_{1,C}(\xi_r)$, for some values of $\xi_r$  within the range of validity of (\ref{eq:Druyvesteyn-weak_absorption})\@.  One can observe that the ratio $Z^\prime_{x,C}(\xi_r)/Z^\prime_{1,C}(\xi_r)$ varies rapidly around $x = 1$ with a slope growing vary fast with $\xi_r$.

The plots show that not only the value of $F_x(v_\parallel)$ diminishes as $x$ increases, as can be surmised in Fig. \ref{fig:druyvesteyn-dist}, but its derivative is also greatly reduced.  Consequently, the damping of long-wavelength Langmuir waves in super-Maxwellian plasmas is greatly curbed by the lack of high-energy electrons and by the flatter profile of the VDF at these speeds.  This effect is observed in the top right panel of Fig. \ref{fig:DRU1:Tischmann-2024-1}.

The opposite effect happens with sub-Maxwellian plasmas.  Fig. \ref{fig:DRU1:Druyvesteyn_Zp_ratios} shows that $F^\prime_x(v_\parallel)$ quickly increases as $x$ is reduced and, consequently, the damping rate rises very fast, as one can observe in the right bottom panel of Fig. \ref{fig:DRU1:Tischmann-2024-1}\@.
We can thus summarize this analysis by remarking that the analytical expression for $\gamma_x(k_\parallel)$ shows that Langmuir waves are more intensely damped at a given wavenumber the lower the parameter $x$.
%
The excess of electrons able to tap the energy transported by Langmuir waves in a sub-Maxwellian plasma, combined with the higher flux velocity that the waves possess in these systems, hint at a very different scenario of evolution of the plasma turbulence with the subsequent associated nonlinear processes, such as particle energization and radiation emission at the fundamental and harmonics of the local electron gyrofrequency.

As the wavenumber of the waves increases, the resonant velocities approach the core of the distribution and expression (\ref{eq:Druyvesteyn-weak_absorption}) becomes invalid.  Plots of $F_x^\prime(v_\parallel)$ show that, irrespective to the value of $x$, the derivative of the VDF increases fast and substantially when $|v_\parallel/\theta| \lesssim 2$, which means that when $\xi_r \simeq 1$ not only the approximation is no longer valid, but also that the damping rate can be of the same order of the real part.  Fig. \ref{fig:DRU1:Tischmann-2024-1} shows that in a thermal plasma, the expression for $\gamma$ is accurate only for $\lambda_{De}k_\parallel \lesssim 0.6$\@.  Assuming the same condition for all $x$, but replacing $\lambda_{De} \to \lambda_{De,x}$,  given by (\ref{eq:DRU1:Effective-Debye_length}),
we conclude that the analytic expression is valid only for
\begin{displaymath}
\lambda_{De} k_\parallel \lesssim \frac{0.6}{g_x}.
\end{displaymath}
Hence, the critical value of $\lambda_{De} k_\parallel$ for the validity of the weak resonance expressions increases with $x$, as can be observed in all panels of Fig. \ref{fig:DRU1:Tischmann-2024-1}.

%

\section{Summary}
With the present paper we have introduced the concept of a Druyvesteyn plasma, wherein the plasma species are characterized by generalized Druyvesteyn distributions. The latter exhibit not only high-energy tails for a Druyvesteyn parameter $x<1$, but also low-energy flat-tops for $x>1$, which are observed in astrophysical plasmas 
(see, e.g., the respective overviews provided in \citet{Lazar-Fichtner-2021} and \citet{Stasiewicz-2024}).


Given the recent interest in electrostatic waves in astrophysical plasmas, with this first paper we have provided the dispersion relation for high-frequency longitudinal waves propagating along an ambient, homogeneous magnetic field in a Druyvesteyn plasma in terms of the (derivative of) the newly defined Druyvesteyn dispersion function and have investigated Langmuir waves. 
For these electron waves we have computed the dispersion curves and damping rates for both flat-top distributions and those with high-energy tails. The first case reproduces and generalizes the results obtained previously by \citet{Amemiya-2012} for the original Druyvesteyn distribution and the second case is qualitatively comparable to findings obtained with different suprathermal distribution, namely regularized Kappa distributions. This findings particularly comprise the occurrence of anomalous dispersion implying negative group velocity. The fact that all results that have been computed fully-numerically from the general dispersion relation and semi-analytically with the dispersion relation expressed in terms of
the Druyvesteyn dispersion function are consistent corroborates the validity of the latter. 

Furthermore, analytical expressions for both the dispersion relation and damping rate of high-frequency (Langmuir) longitudinal waves propagating in a Druyvesteyn plasma were obtained from the usual weak-damping approximation.  As it happens in the Maxwellian case, the approximate expressions are accurate in the low-wavenumber limit for all values of the parameter $x$, but the spectral range where the approximations are good increases with $x$\@.  

In all cases, the approximate expression for the damping rate predicts weak absorption of Langmuir waves throughout the spectral range, whereas the full numerical solution shows that absorption (apparently) always increases with wavenumber.  Moreover, the approximate dispersion relation fails to reproduce the anomalous behavior $(\partial\omega_r/\partial k < 0)$ displayed by the case $x < 1$ for sufficiently short wavelengths.  

With the introduction of the Druyvesteyn plasma we have provided a new, versatile tool for the quantitative treatment of linear waves in non-Maxwellian plasmas that generalizes previous work and supplements existing alternatives like the so-called Kappa plasmas. 
Following the analysis of Langmuir waves in the present paper, the next study suggests itself, namely that of ion-acoustic waves, which are of interest for the solar wind \citep{Vidal-Luengo-etal-2025}, for planetary environments \citep{Morsi-etal-2024}, for the interstellar medium \citep{Gao-etal-2024}, and for astrophysical dusty plasmas \citep{Lazar-etal-2018}.

\appendix{}

\section{Properties and evaluation of the Druyvesteyn dispersion function}\label{Zx-Properties}

%

Several properties of the derivative $Z_{x}^{\prime}\left(\xi\right)$
of the Druyvesteyn dispersion function can be obtained
from (\ref{eq:DRU1:Druyvesteyn_dispersion_function-derivative}),
some of which will be derived here.

\subsection{Value at origin}

The value of $Z_{x}^{\prime}\left(\xi=0\right)$ can be obtained directly
from (\ref{eq:DRU1:Druyvesteyn_dispersion_function-derivative}),
which gives
\[
Z_{x}^{\prime}\left(0\right)=-\frac{x}{\Gamma\left(\nicefrac{3}{2x}\right)}\int_{-\infty}^{\infty}du\,e^{-u^{2x}}.
\]

The remaining integral is a particular case of
\[
I_{n}=\int_{-\infty}^{\infty}du\,u^{2n}e^{-u^{2x}}\quad\left(n=0,1,2,\dots\right).
\]
Recalling that the exponential must be evaluated as $e^{-u^{2x}}=\exp\left[-\left(u^{2}\right)^{x}\right]$,
the integrand is even. Then, upon defining the new integration variable
$y=u^{2x}$,
\[
I_{n}=x^{-1}\int_{0}^{\infty}dy\,y^{\left(n+1/2\right)x^{-1}-1}e^{-y}.
\]
Using now the definition of the gamma function \citep{AskeyRoy-Full-NIST10},
one finally obtains
\begin{equation}
\int_{-\infty}^{\infty}du\,u^{2n}e^{-u^{2x}}=x^{-1}\Gamma\left(\left(n+\frac{1}{2}\right)x^{-1}\right).\label{eq:I_n}
\end{equation}

Therefore,
\[
Z_{x}^{\prime}\left(0\right)=-\frac{\Gamma\left(\nicefrac{1}{2x}\right)}{\Gamma\left(\nicefrac{3}{2x}\right)}.
\]
When $x=1$, this result reduces to $Z_{1}^{\prime}\left(0\right)=Z^{\prime}\left(0\right)=-2$.

\subsection{Analytic continuation}

The function $Z_{x}^{\prime}\left(\xi\right)$ can be directly evaluated
from (\ref{eq:DRU1:Druyvesteyn_dispersion_function-derivative}) using
a numerical quadrature routine. However, when $\Im\xi\leqslant0$
one needs the analytic continuation, which is provided by the Landau
prescription of deforming the integration contour in such a way that
it always remains below the pole at $u=\xi_{r}+i\xi_{i}$ \citep{KrallTrivelpiece86}.

Therefore, upon using the Sokhotski-Plemelj theorem, we can write
\begin{equation}
Z_{x}^{\prime}\left(\xi\right)=Z_{x,NC}^{\prime}\left(\xi\right)+iZ_{x,C}^{\prime}\left(\xi\right),\label{eq:DRU1:DDFP-Analytic_continuation}
\end{equation}
where
\begin{eqnarray*}
Z_{x,NC}^{\prime}\left(\xi\right) & = & -\frac{x}{\Gamma\left(\nicefrac{3}{2x}\right)}\int_{-\infty}^{\infty}du\,\frac{ue^{-u^{2x}}}{u-\xi},\\
Z_{x,C}^{\prime}\left(\xi\right) & = & -2\pi\epsilon\frac{x\xi e^{-\xi^{2x}}}{\Gamma\left(\nicefrac{3}{2x}\right)},
\end{eqnarray*}
and where $\epsilon = 0$ (for $\xi_i > 0$), $\nicefrac{1}{2}$ (for $\xi_i = 0$), or 1 (for $\xi_i < 0$)\@.  
In (\ref{eq:DRU1:DDFP-Analytic_continuation}), the acronym ``$NC$''
stands for \emph{non-continued}, whereas ``$C$'' means \emph{continued}.

\subsection{Asymptotic expansion}

When $\left|\xi\right|$ is sufficiently large, we can derive an asymptotic
expansion for $Z_{x}^{\prime}\left(\xi\right)$\@. Starting from
the geometric progression, we can write in (\ref{eq:DRU1:Druyvesteyn_dispersion_function-derivative}),
\[
\frac{1}{1-u^{2}/\xi^{2}}=\sum_{k=0}^{N}\frac{u^{2k}}{\xi^{2k}}+\frac{1}{\xi^{2\left(N+1\right)}}\frac{u^{2\left(N+1\right)}}{1-u^{2}/\xi^{2}},
\]
for a given $N>0$\@. The last term on the RHS can be used to evaluate
an error bound for the asymptotic formula.

Leaving the error analysis for a future publication, we will simply
assume that $\left|\xi\right|$ is sufficiently large so that a finite
number of terms in the sum are enough to obtain a target accuracy
and approximate the non-continued part of the function as
\[
Z_{x,NC}^{\prime}\left(\xi\right)\simeq\frac{2x}{\Gamma\left(\nicefrac{3}{2x}\right)}\sum_{k=0}\frac{1}{\xi^{2\left(k+1\right)}}\int_{0}^{\infty}du\,u^{2\left(k+1\right)}e^{-u^{2x}}.
\]
Employing again formula (\ref{eq:I_n}), we obtain the asymptotic
expansion
\begin{equation}
Z_{x,NC}^{\prime}\left(\xi\right)\simeq\frac{1}{\Gamma\left(\nicefrac{3}{2x}\right)}\sum_{k=0}\frac{\Gamma\left[\left(k+\nicefrac{3}{2}\right)x^{-1}\right]}{\xi^{2\left(k+1\right)}},\label{eq:DRU1:DDF-asymptotic}
\end{equation}
which must be complemented with $Z_{x,C}^{\prime}\left(\xi\right)$, given by (\ref{eq:DRU1:DDFP-Analytic_continuation}).

\subsection{Mellin transform and series representations}

The Mellin transform method has been successfully applied to obtain
computable representations for the plasma dispersion function resulting
from standard Kappa distributions \citep{GaelzerZiebell16/02,Gaelzer+16/06} as well as the regularized Kappa distribution \citep{Gaelzer-etal-2024}\@.
Here, we will derive representations for $Z_{x}^{\prime}\left(\xi\right)$
in terms of the Fox $H$-function introduced in Appendix \ref{sec:Fox_function}
and subsequently derive series representations that are adequate for
numerical computation.

Returning to (\ref{eq:DRU1:Druyvesteyn_dispersion_function-derivative}),
we can define the new integration variable $t=u^{2}$ and, by means of
a simple algebraic manipulation, write
\[
Z_{x}^{\prime}\left(\xi\right)=\frac{x}{\Gamma\left(\nicefrac{3}{2x}\right)\xi^{2}}\int_{0}^{\infty}dt\,\frac{t^{1/2}e^{-t^{x}}}{1-\xi^{-2}t}.
\]


Identifying the integrand above with the representations in section \ref{app:a4} 
we obtain
\[
Z_{x}^{\prime}\left(\xi\right)=\frac{\xi^{-2}}{\Gamma\left(\nicefrac{3}{2x}\right)}\int_{0}^{\infty}dt\,t^{1/2}H_{1,1}^{1,1}\left[-\xi^{-2}t\left|{\left(0,1\right)\atop \left(0,1\right)}\right.\right]H_{0,1}^{1,0}\left[t\left|{-\atop \left(0,x^{-1}\right)}\right.\right].
\]
Using now property (\ref{eq:FoxH1:Mellin_transf-2H}), we obtain our
first representation,
\begin{equation}
Z_{x}^{\prime}\left(\xi\right)=\frac{-i\xi^{-1}}{\Gamma\left(\nicefrac{3}{2x}\right)}H_{1,2}^{2,1}\left[-\xi^{2}\left|{\left(\nicefrac{1}{2},1\right)\atop \left(x^{-1},x^{-1}\right),\left(\nicefrac{1}{2},1\right)}\right.\right].\label{eq:DRU1:ZpH-1}
\end{equation}

However, an alternative and equivalent result can by derived if we
insert the identity
\[
\left(-1\right)^{-s}=\frac{\pi}{\Gamma\left(\nicefrac{1}{2}+s\right)\Gamma\left(\nicefrac{1}{2}-s\right)}+i\frac{\pi}{\Gamma\left(s\right)\Gamma\left(1-s\right)}
\]
into the explicit integral expression of (\ref{eq:DRU1:ZpH-1}) and
then perform the possible simplifications. This procedure leaves us
with the result
\begin{equation}
Z_{x}^{\prime}\left(\xi\right)=\frac{\pi}{\Gamma\left(\nicefrac{3}{2x}\right)}\left\{ H_{2,3}^{2,1}\left[\xi^{2}\left|{\left(0,1\right),\left(-\nicefrac{1}{2},1\right)\atop \left(x^{-1}/2,x^{-1}\right),\left(0,1\right),\left(-\nicefrac{1}{2},1\right)}\right.\right]-ix\xi e^{-\xi^{2x}}\right\} ,\label{eq:DRU1:ZpH-2}
\end{equation}
where we have used the translation property (\ref{eq:FoxH1:Trans_prop-2})
and once again representation (\ref{eq:FoxH1:Exp-H}).

The representation (\ref{eq:DRU1:ZpH-2}) is particularly important
because when $x=1$, it reduces to
\begin{equation}
Z_{1}^{\prime}\left(\xi\right)=Z^{\prime}\left(\xi\right)=-2\left[M\left({1\atop \nicefrac{1}{2}};-\xi^{2}\right)+i\sqrt{\pi}\xi e^{-\xi^{2}}\right],\label{eq:Zp-M}
\end{equation}
which was obtained using (\ref{eq:FoxH1:G_H_representation}) and
(\ref{eq:MeiG1:1F1-G})\@. This is one of the known representations
for the derivative of the Fried \& Conte function in terms of the
confluent hypergeometric function \citep{Peratt84/03}\@.


From (\ref{eq:DRU1:ZpH-2}) one can derive series expansions for the
$H$-function. For particular values of the parameter $x$, power
series of the argument $\xi^{2}$ are possible, but for other values
power-logarithmic series must be derived. The conditions for either
case are discussed at length in \citet{KilbasSaigo04} and will be
briefly reproduced here for the particular $H$-function in representation
(\ref{eq:DRU1:ZpH-2}), which will be denoted as $H\left(z\right)$,
for brevity.

According to (\ref{eq:FoxH1:Fox_H-function}),

\begin{equation}
H\left(z\right) = \frac{1}{2\pi i}\int_{L}\mathcal{H}\left(s\right)z^{-s} ds,
\text{ where }
\mathcal{H}\left(s\right) = \frac{\Gamma\left(s\right)\Gamma\left(x^{-1}/2+x^{-1}s\right)\Gamma\left(1-s\right)}{\Gamma\left(\nicefrac{3}{2}-s\right)\Gamma\left(-\nicefrac{1}{2}+s\right)}.
\label{eq:DRU1:ZpH-3} \\
\end{equation}
Let us consider the conditions for a power series expansion of $H\left(z\right)$\@.
According to the discussion in section \ref{sec:FoxH1:Power_series},
the poles of the function $\Gamma\left(s\right)$ occur at $s_{1\ell^\prime}=-\ell^{\prime}$
$\left(\ell^{\prime}=0,1,2,\dots\right)$, whereas the poles of $\Gamma\left(x^{-1}/2+x^{-1}s\right)$
occur at $s_{2\ell}=-\ell x-\nicefrac{1}{2}$ $\left(\ell=0,1,2,\dots\right)$\@.
Hence, the function $H\left(z\right)$ can be represented by power
series if $s_{1\ell^{\prime}}\neq s_{2\ell}$ for any pair $\left(\ell^{\prime},\ell\right)$.

Therefore, if we define the rational parameter
\[
x_{mn}= \frac{2m-1}{2n}\;\left(m,n=1,2,\dots\right),
\]
which corresponds to any rational number composed by an odd positive
integer over an even integer, the function $H\left(z\right)$ can
be expanded according to (\ref{eq:FoxH1:Power_series-1}) whenever
$x\neq x_{mn}$\@. In this case, we obtain the power series representation
\begin{equation}
Z_{x}^{\prime}\left(\xi\right) = -\frac{\pi}{\Gamma\left(\nicefrac{3}{2x}\right)}\left[x\xi\sum_{\ell=0}^{\infty}\tan\left(\pi\ell x\right)\frac{\left(-\xi^{2x}\right)^{\ell}}{\ell!}
+ \frac{1}{\pi}\sum_{\ell=0}^{\infty}\Gamma\left(\frac{x^{-1}}{2}-\ell x^{-1}\right)\xi^{2\ell}+ix\xi e^{-\xi^{2x}}\right].
\label{eq:DRU1:Zp-Power_series}
\end{equation}%
One can easily verify that $Z_{1}^{\prime}\left(\xi\right)=Z^{\prime}\left(\xi\right)$,
because formula (\ref{eq:Zp-M}) is reproduced.

Let us now consider the case $x=x_{mn}$\@. Let us also define the
set $\mathcal{L}_{d}=\left\{ \left(\ell^{\prime},\ell\right)\right\} $
composed by all pairs of indices that satisfy the condition $\ell=\left(\ell^{\prime}-\nicefrac{1}{2}\right)x^{-1}\in\mathbb{N}^{+}$,
and the sets $\mathcal{L}_{s1}=\left\{ \ell^{\prime}\right\} \setminus\mathcal{L}_{d}$
and $\mathcal{L}_{s2}=\left\{ \ell\right\} \setminus\mathcal{L}_{d}$,
respectively composed by the indices $\ell^{\prime}$ and $\ell$
that are not contained in any pair of $\mathcal{L}_{d}$\@. Whenever
a pair $\left(\ell^{\prime},\ell\right)$ is an element of $\mathcal{L}_{d}$,
the corresponding pole in the integrand of $H\left(z\right)$ is of
second order and its contribution to the integral must be evaluated
separately with the residue theorem.


Therefore, we can evaluate the function $H\left(z\right)$ in (\ref{eq:DRU1:ZpH-3})
as
\[
H\left(z\right)=\sum_{\ell^{\prime}\in\mathcal{L}_{s1}}
\mathrm{Res}\left( \mathcal{H}_{2,3}^{2,1}\left(s\right)z^{-s}; s_{1\ell^\prime}
\right)
+
\sum_{\ell\in\mathcal{L}_{s2}} \mathrm{Res}\left( \mathcal{H}_{2,3}^{2,1}\left(s\right)z^{-s}; s_{2\ell} \right)
+
\sum_{\ell^{\prime}\in\mathcal{L}_{d}}
\mathrm{Res}\left( \mathcal{H}_{2,3}^{2,1}\left(s\right)z^{-s}; s_{1\ell^\prime}
\right).
\]

In this way, after a fair amount of algebra, we obtain the following
representation,
\begin{eqnarray}
Z_x^\prime \left(\xi\right) & = & -\frac{\pi}{\Gamma\left( \nicefrac{3}{2x} \right)}
\left\{ ix\xi e^{-\xi^{2x}} + \frac{1}{\pi} \sum_{\ell^\prime \in \mathcal{L}_{s1}}
\Gamma\left( \frac{x^{-1}}{2} - \ell^\prime x^{-1} \right) \xi^{2\ell^\prime}
\right.
\nonumber \\
 & & \left.
+x\xi \sum_{\ell\in\mathcal{L}_{s2}} \tan\left(\ell\pi x\right)
\frac{\left( -\xi^{2x} \right)^\ell}{\ell!}
- \frac{2}{\pi}x\xi \sum_{\ell\in\mathcal{L}_d}
\left[ \ln\xi - \frac{1}{2}x^{-1} \psi\left( \ell + 1 \right) \right]
\frac{\left( -\xi^{2x} \right)^\ell}{\ell!}
\right\},
\label{eq:DRU1:Zp-Power-log_series}
\end{eqnarray}
which contains a power-logarithmic expansion, where $\psi\left(z\right)=\Gamma^{\prime}\left(z\right)/\Gamma\left(z\right)$
is the digamma function \citep{AskeyRoy-Full-NIST10}.


\section{The Fox $H$- and the Meijer $G$-functions}\label{sec:Fox_function}

The Fox $H$-function is that function whose Mellin transform can
be expressed as a ratio of certain products of gamma functions. Consequently,
its definition is given by the Mellin-Barnes contour integral \citep{KilbasSaigo04,Mathai+09}
\begin{equation}
H_{p,q}^{m,n}\left[z\left|{\left(a_{p},\alpha_{p}\right)\atop \left(b_{q},\beta_{q}\right)}\right.\right]=\frac{1}{2\pi i}\int_{L}\mathcal{H}_{p,q}^{m,n}\left(s\right)z^{-s}ds,\label{eq:FoxH1:Fox_H-function}
\end{equation}
where 
\[
\mathcal{H}_{p,q}^{m,n}\left(s\right)=\frac{\prod_{j=1}^{m}\Gamma\left(b_{j}+\beta_{j}s\right)\prod_{i=1}^{n}\Gamma\left(1-a_{i}-\alpha_{i}s\right)}{\prod_{j=m+1}^{q}\Gamma\left(1-b_{j}-\beta_{j}s\right)\prod_{i=n+1}^{p}\Gamma\left(a_{i}+\alpha_{i}s\right)}.
\]

In (\ref{eq:FoxH1:Fox_H-function}), $m,n,p,q\in\mathbb{N}$, $0\leqslant m\leqslant q$,
$0\leqslant n\leqslant p$, $\left\{ \alpha_{j},\beta_{j}\right\} \in\mathbb{R}^{+}$,
and $\left\{ a_{j},b_{j}\right\} \in\mathbb{C}$\@. If $m+1>q$ or
$n+1>p$, the product is replaced by one. The notation is such that
$\left(a_{p},\alpha_{p}\right)=\left(a_{1},\alpha_{1}\right),\dots,\left(a_{p},\alpha_{p}\right)$
and $\left(b_{q},\beta_{q}\right)=\left(b_{1},\beta_{1}\right),\dots,\left(b_{q},\beta_{q}\right)$\@.
It is assumed that the poles 
\[
s_{j\ell}=-\left(\frac{b_{j}+\ell}{\beta_{j}}\right),\quad\left(j=1,\dots,m;\:\ell\in\mathbb{N}\right)
\]
of the functions $\Gamma\left(b_{j}+\beta_{j}s\right)$ do not coincide
with the poles 
\[
\sigma_{ik}=\left(\frac{1-a_{i}+k}{\alpha_{i}}\right),\quad\left(i=1,\dots,n;\:k\in\mathbb{N}\right)
\]
of the functions $\Gamma\left(1-a_{i}-\alpha_{i}s\right)$; that is,
$\alpha_{i}\left(b_{j}+\ell\right)\neq\beta_{j}\left(a_{i}-k-1\right)$,
($j=1,\dots,m$, $i=1,\dots,n$, $\ell,k\in\mathbb{N}$).

The integration contour $L$ in (\ref{eq:FoxH1:Fox_H-function}) is
deformed in such a way that it separates all the poles $\left\{ s_{j\ell}\right\} $
to the left and all the poles $\left\{ s_{ik}\right\} $ to the right
of $L$\@. This is accomplished by choosing one of a total of 3 different
contour types for $L$, which will result in different representations
for the $H$-function \citep{KilbasSaigo04}.

The $H$-function has remarkable mathematical properties. For instance,
all generalized hypergeometric functions are particular cases,
but it also contains functions that can not be expanded in power series
anywhere, such as functions with logarithmic singularities. Below,
we present some relevant mathematical properties.

\subsection{Translation property}

The following property is relevant,
\begin{equation}
z^{\sigma}H_{p,q}^{m,n}\left[z\left|{\left(a_{p},\alpha_{p}\right)\atop \left(b_{q},\beta_{q}\right)}\right.\right]=H_{p,q}^{m,n}\left[z\left|{\left(a_{p}+\sigma\alpha_{p},\alpha_{p}\right)\atop \left(b_{q}+\sigma\beta_{q},\beta_{q}\right)}\right.\right],\quad\left(\sigma\in\mathbb{C}\right).\label{eq:FoxH1:Trans_prop-2}
\end{equation}

\subsection{Mellin transform of two $H$-functions}

The $H$-function has the remarkable property that the Mellin transform
of the product of two $H$-functions is itself a Fox function, 
\begin{eqnarray}
\lefteqn{\hspace*{-1.5cm}\int_{0}^{\infty}dx\,x^{s-1}H_{p,q}^{m,n}\left[zx^{\sigma}\left|{\left(a_{p},\alpha_{p}\right)\atop \left(b_{q},\beta_{q}\right)}\right.\right]H_{P,Q}^{M,N}\left[wx\left|{\left(c_{P},\gamma_{P}\right)\atop \left(d_{Q},\delta_{Q}\right)}\right.\right]}\nonumber \\
 & \hspace*{1.5cm}= & w^{-s}H_{p+Q,q+P}^{m+N,n+M}\left[\frac{z}{w^{\sigma}}\left|{\left(a_{i},\alpha_{i}\right)_{1,n},\left(1-d_{Q}-s\delta_{Q},\sigma\delta_{Q}\right),\left(a_{i},\alpha_{i}\right)_{n+1,p}\atop \left(b_{j},\beta_{j}\right)_{1,m},\left(1-c_{P}-s\gamma_{P},\sigma\gamma_{P}\right),\left(b_{j},\beta_{j}\right)_{m+1,q}}\right.\right].\label{eq:FoxH1:Mellin_transf-2H}
\end{eqnarray}

\subsection{The Meijer $G$-function as a particular case}

In (\ref{eq:FoxH1:Fox_H-function}), if all $\left\{ \alpha_{p}\right\} $
and $\left\{ \beta_{q}\right\} $ are unitary, one obtains the Meijer
$G$-function 
\begin{equation}
G_{p,q}^{m,n}\left[z\left|{\left(a_{p}\right)\atop \left(b_{q}\right)}\right.\right]=H_{p,q}^{m,n}\left[z\left|{\left(a_{p},1\right)\atop \left(b_{q},1\right)}\right.\right]=\frac{1}{2\pi i}\int_{L}\mathcal{G}_{p,q}^{m,n}\left(s\right)z^{-s}ds,\label{eq:FoxH1:G_H_representation}
\end{equation}
where 
\[
\mathcal{G}_{p,q}^{m,n}\left(s\right)=\frac{\prod_{j=1}^{m}\Gamma\left(b_{j}+s\right)\prod_{i=1}^{n}\Gamma\left(1-a_{i}-s\right)}{\prod_{j=m+1}^{q}\Gamma\left(1-b_{j}-s\right)\prod_{i=n+1}^{p}\Gamma\left(a_{i}+s\right)}.
\]

Properties of the $G$-function were presented in \citet{GaelzerZiebell16/02}
and \citet{Gaelzer+16/06} and in the cited literature.

\subsection{Representations of elementary and special functions}
\label{app:a4}

Given a function $f\left(z\right)$ $\left(z\in\mathbb{C}\right)$,
its representation in terms of a $H$- or a $G$-function is usually
obtained by applying the Mellin transform 
\[
F\left(s\right)=\mathcal{M}\left\{ f\right\} =\int_{0}^{\infty}z^{s-1}f\left(z\right)dz
\]
and then by identifying $F\left(s\right)$ with the integrands either
in (\ref{eq:FoxH1:Fox_H-function}) or in (\ref{eq:FoxH1:G_H_representation}).


Here, we employ the representations
\begin{eqnarray}
\frac{1}{\left(1+az^{h}\right)^{\nu}} & = & \frac{1}{\Gamma\left(\nu\right)}H_{1,1}^{1,1}\left[az^{h}\left|{\left(1-\nu,1\right)\atop \left(0,1\right)}\right.\right]\\
z^{b/\beta}\exp\left(-z^{1/\beta}\right) & = & \beta H_{0,1}^{1,0}\left[z\left|{-\atop \left(b,\beta\right)}\right.\right]\label{eq:FoxH1:Exp-H}\\
M\left({a\atop b};z\right) & = & \frac{\Gamma\left(b\right)}{\Gamma\left(a\right)}G_{1,2}^{1,1}\left[-z\left|{1-a\atop 0,1-b}\right.\right].\label{eq:MeiG1:1F1-G}
\end{eqnarray}

The representation (\ref{eq:FoxH1:Exp-H}) can be obtained using identity
(\ref{eq:I_n})\@. In (\ref{eq:MeiG1:1F1-G}), $M\left(z\right)$
is the Kummer confluent hypergeometric function \citep{Daalhuis-Full-NIST10a}.

\subsection{Power series expansion}\label{sec:FoxH1:Power_series}

In (\ref{eq:FoxH1:Fox_H-function}), if all the poles of the gamma
functions $\Gamma\left(b_{j}+\beta_{j}s\right)$ $\left(j=1,\dots,m\right)$
are simple, \emph{i.e.}, if
\begin{equation}
\beta_{j^{\prime}}\left(b_{j}+k\right)\neq\beta_{j}\left(b_{j^{\prime}}+\ell\right),\quad\left(j,j^{\prime}=1,\dots,m,\,j\neq j^{\prime};\,k,\ell\in\mathbb{N}\right),\label{eq:Simple_pole-conditions}
\end{equation}
then the residue theorem can be applied to the integration contour
$L$ in (\ref{eq:FoxH1:Fox_H-function}) that loops around all poles
$\left\{ s_{j\ell}\right\} $ $\left(j=1,\dots,m\right)$, and the
$H$-function can be evaluated as 
\begin{displaymath}
H_{p,q}^{m,n}\left[z\left|{\left(a_{p},\alpha_{p}\right)\atop \left(b_{q},\beta_{q}\right)}\right.\right]=\sum_{j=1}^{m}\sum_{\ell=0}^{\infty}
\mathrm{Res}\left( \mathcal{H}_{p,q}^{m,n}\left(s\right)z^{-s}; s_{j\ell} \right),
\end{displaymath}
which provides the power series expansion 
\begin{equation}
H_{p,q}^{m,n}\left[z\left|{\left(a_{p},\alpha_{p}\right)\atop \left(b_{q},\beta_{q}\right)}\right.\right]=\sum_{h=1}^{m}\sum_{\ell=0}^{\infty}\frac{{\displaystyle \prod_{j=1 \atop
j\neq h
}^{m}\Gamma\left(b_{j}-\left(b_{h}+\ell\right)\frac{\beta_{j}}{\beta_{h}}\right)\prod_{j=1}^{n}\Gamma\left(1-a_{j}+\left(b_{h}+\ell\right)\frac{\alpha_{j}}{\beta_{h}}\right)}}{{\displaystyle \prod_{j=n+1}^{p}\Gamma\left(a_{j}-\left(b_{h}+\ell\right)\frac{\alpha_{j}}{\beta_{h}}\right)\prod_{j=m+1}^{q}\Gamma\left(1-b_{j}+\left(b_{h}+\ell\right)\frac{\beta_{j}}{\beta_{h}}\right)}}\frac{\left(-\right)^{\ell}z^{\left(b_{h}+\ell\right)/\beta_{h}}}{\beta_{h}\ell!}.\label{eq:FoxH1:Power_series-1}
\end{equation}

Conversely, if one or more of the conditions (\ref{eq:Simple_pole-conditions})
are not satisfied, then those poles are of higher order and the residue
theorem will provide power-logarithm expansions for the $H$-function.

\begin{acknowledgments}
This work was carried out within a bilateral research project funded by the
\textit{Deutsche Forschungsgemeinschaft (DFG, FI 706/31-1)} and the 
\textit{Belgian FWO-Vlaanderen (G002522N)}. 
R.G. acknowledges support provided by Conselho Nacional de Desenvolvimento Científico e Tecnológico (CNPq), Grant No. 313330/2021-2\@.
M.L. acknowledges support in the framework of the project 4000145223 SIDC Data Exploitation (SIDEX2), ESA Prodex.
H.F. thanks M.J.\ P\"uschel for discussions on anomalous dispersion during a mini-workshop on Astro and Space Plasmas at DIFFER, Eindhoven.
\end{acknowledgments}


\bibliography{druyvesteyn_r2}



\end{document}